\documentclass[final,5p,times,twocolumn]{elsarticle}
\graphicspath{{fig/}}

\usepackage{amssymb}
\usepackage{lipsum}
\usepackage{amsmath}
\usepackage{hyperref}
\usepackage{booktabs}
\biboptions{numbers,sort&compress}

\journal{Nuclear Physics A}

\begin{document}

\begin{frontmatter}

\title{Performance of a liquid nitrogen cryostat setup for the study of nuclear recoils in undoped CsI crystals}

\author[a]{K. Ding}
\author[a]{J. Liu}
\author[a]{Y. Yang}
\author[b]{K. Scholberg}
\author[c,d]{D.M. Markoff}

\affiliation[a]{Physics Department, University of South Dakota, Vermillion, SD, 57069, USA}
\affiliation[b]{Department of Physics, Duke University, Durham, NC, 27708, USA}
\affiliation[c]{Triangle Universities Nuclear Laboratory, Durham, NC, 27708, USA}
\affiliation[d]{Department of Mathematics and Physics, North Carolina Central University, Durham, NC, 27707, USA}

\begin{abstract}
There is a global trend to increase the light yield of CsI scintillators used in neutrino and dark matter detection by operating undoped crystals at cryogenic temperatures. However, high light yield alone is not sufficient to guarantee a low-energy threshold. The response of undoped crystals to nuclear recoils at cryogenic temperatures is equally important. A liquid nitrogen-based cryostat was developed to measure the nuclear quenching factor of a small undoped CsI crystal using monoenergetic neutron beams at the Triangle Universities Nuclear Laboratory (TUNL). To avoid neutron scattering in high-$Z$ materials, these materials were intentionally reduced around the crystal. The structure and performance of the cryostat are described in detail. Using this cryostat, a light yield of $33.4 \pm 1.7$ photoelectrons per keV electron-equivalent (PE/keV$_\text{ee}$) was observed at 5.9 keV$_\text{ee}$, enabling the measurement of nuclear quenching factors at very low energies. The results of the quenching factor measurement will be reported in a subsequent paper.

Non-negligible negative overshoot was observed in the tails of the observed light pulses. The origin of this issue and the correction procedure are described in detail. This information may be useful for others who encounter similar technical challenges.
\end{abstract}


\end{frontmatter}

\section{Introduction}
\label{s:intro}

Undoped cesium iodide (CsI) crystals at around liquid nitrogen temperature are very bright scintillators~\cite{Nishimura95, Amsler02, Sailer12}. When combined with photo-sensors working at cryogenic temperatures, the light yield of these crystals can be at least twice as high as that of Tl- or Na-doped CsI operated at room temperature~\cite{coherent17}. The light yields of undoped CsI crystals from SICCAS, AMCRYS, and OKEN, with various sizes, from 13 keV to 2.6 MeV, and directly coupled with PMTs or SiPMs at liquid nitrogen temperature were systematically examined~\cite{csi, csi20, ding20e, ding20, ding22}. The highest light yield obtained was 43.0 PE/keV$_\text{ee}$, achieved with a small crystal coupled to two SensL J-series SiPMs~\cite{ding22}.

With such high light yields, the energy threshold of this detector can be significantly lower than that of a doped CsI detector. A 14.6~kg Na-doped CsI detector was used by the COHERENT Collaboration to observe $134 \pm 22$ coherent elastic neutrino($\nu$)-nucleus scattering (CEvNS) events with two years of operation at the Spallation Neutron Source (SNS), in the Oak Ridge National Laboratory (ORNL)~\cite{coherent17}. Assuming a similar configuration and exposure, nearly 1350 CEvNS events are expected with a high light yield undoped CsI detector operated at liquid nitrogen temperature. Sensitivities of such a detector for probing non-standard neutrino interactions~\cite{davidson03, bar05, coloma16, coloma17, liao17, papoulias18, denton18, nsi19} and low-mass dark matter particles~\cite{Kobzarev:1966qya, Blinnikov:1982eh, foot91, hodges93, berezhiani96, PhysRevLett.39.165, Izaguirre:2015yja, fayet90} are discussed in detail in Ref.~\cite{csi20, ding20}.

Those sensitivity studies were based on the assumption that undoped crystals at 77~K have a similar nuclear scintillation quenching effect to undoped ones at room temperature. A recent study at 108~K~\cite{CollarQF} confirmed this assumption. However, measurements done with $\alpha$ particles~\cite{clarkQF} revealed strong temperature dependence of the quenching effect. It is therefore important to measure the nuclear quenching effect closer to 77~K.

The nuclear quenching measurement is normally done by putting the material under study in a neutron beam. The energy of the nuclear recoil can then be calculated from the energy of the incident neutron and the angle of the scattered neutron, assuming the neutron only scatters once within the target. High $Z$ material should therefore be avoided around the target to minimize the occurrence of neutron multiple scatterings.

A liquid nitrogen cryostat with low-$Z$ material around a small undoped CsI crystal was developed for the nuclear quenching measurement. Its performance was compared to another cryostat previously used for light yield measurements.

A constant light yield was assumed in those sensitivity calculations. It is known, however, that the light yield of undoped CsI has non-negligible variations at different energy ranges, and it varies with crystals under investigation~\cite{LY_moszynski_energy_2005, LY_gridin_channels_2014, LY_kerisit_computer_2009}. It is hence necessary to verify that the light yields we achieved from 13 keV to 2.6 MeV still hold at lower energies with our crystals purchased from Japan and Ukraine. A $^{55}$Fe source and an $^{241}$Am source were used to achieve this utilizing  5.9 keV x-rays and 60 keV $\gamma$-rays.

\section{Cryostats}
\label{s:setup}
\subsection{Design based on gravity-fed liquid nitrogen dewar}
\begin{figure}[htbp]\centering
  \includegraphics[width=\linewidth]{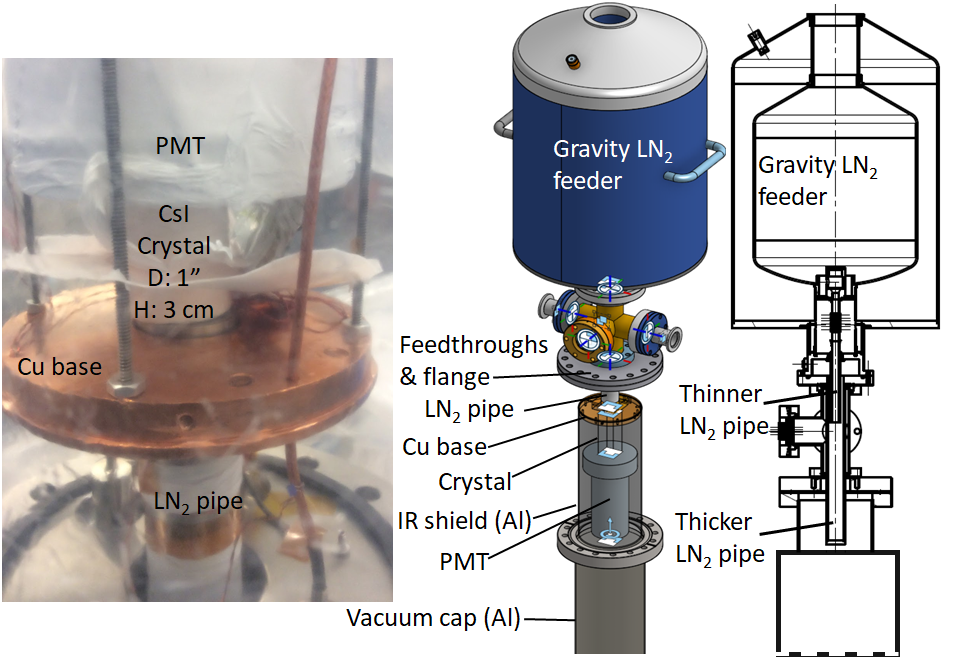}
  \caption{A sketch and photos of the cryostat setup.}
  \label{f:setup}
\end{figure}

Fig.~\ref{f:setup} shows the experimental setup for the measurement of the light yield of an undoped CsI crystal at around 77~K. As seen in the right figure, liquid N$_2$ can drip from a dewar into a hollow pipe that is directly in contact with structures to be cooled. In the detector mounting mode, as shown in the left figure, the dewar can be taken away and this new cryostat can be flipped over and placed into a dry glove box for crystal mounting. The middle figure is the CAD drawing of the cryostat. 

The undoped cylindrical crystal was purchased from AMCRYS~\cite{amcrys}, and had a diameter of 1~inch and a height of 3~cm. All surfaces were mirror polished. The side surface and an end side surface of the crystal were wrapped with multiple layers of Teflon tape to make sure that there was no light leak. A Hamamatsu 3-inch R11065-ASSY PMT was pushed against the other end surface of the crystal by springs to ensure adequate optical contact without optical grease. The springs are held by the copper base. The wrapped end surface of the crystal was pushed against the cooling finger of the cryostat where liquid nitrogen was fed into the LN$_2$ pipe.

To control a reasonable radiation intensity, an $^{241}$Am source and an $^{55}$Fe source were facing opposite towards the center of the crystal. Two layers of aluminum foil were put on the $^{55}$Fe source to further reduce the radiation intensity. The sources were attached to the inner side surface of the aluminum chamber (IR shield) in the cryostat.

To minimize exposure of the crystal to atmospheric moisture, assembly was done in a glove bag flushed with dry nitrogen gas. The relative humidity was kept below 10\% at 22$^{\circ}$C during the assembly process.

The PMT-crystal assembly was capped by an IR shield that was fixed to the copper base by three screws. The cryostat was then sealed by the vacuum cap to a 6-inch ConFlat (CF) flange. A fluorocarbon CF gasket was put in between for multiple operations. The inner diameter of the cryostat was $\sim 10$~cm. Vacuum-welded to the flange were two BNC, two SHV, one 19-pin electronic feedthroughs.

After all cables were fixed beneath it, the top flange was closed. The chamber was then pumped with a Pfeiffer Vacuum HiCube 80 Eco to $\sim 1\times {10}^{-5}$~mbar. The feeder was then filled with LN2 to cool the LN$_2$ pipe, then further cool everything inside. The pump was on all the time until the end of the experiment.

A few Heraeus C~220 platinum resistance temperature sensors were used to monitor the cooling process. They were attached to the side surface of the crystal, the PMT, and the top flange to obtain the temperature profile of the long chamber. A Raspberry Pi 2 computer with custom software~\cite{cravis} was used to read out the sensors. The cooling process to 77~K took about 10 hours. Most measurements, however, were taken after about an additional 2 hours of waiting to let the system reach thermal equilibrium. 

The PMT was powered by a CAEN N1470A high voltage power supply in a NIM crate. The signals were fed into a CAEN DT5751 waveform digitizer, which had a 1~GHz sampling rate, a 1~V dynamic range and a 10 bit resolution. WaveDump~\cite{wavedump}, a free software provided by CAEN, was used for data recording. The recorded binary data files were converted to CERN ROOT files~\cite{root} for analysis by a custom-developed software~\cite{towards}. 

\subsection{Design based on open liquid nitrogen dewar}
\begin{figure}[htbp]\centering
  \includegraphics[width=\linewidth]{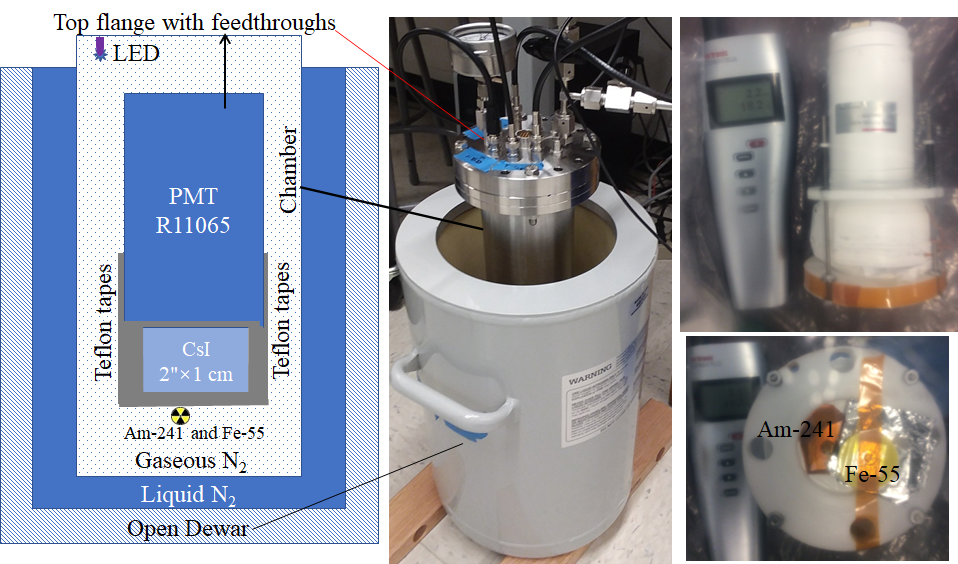}
  \caption{A sketch and photos of the cryostat setup.}
  \label{f:old}
\end{figure}

It is possible to measure the temperature of the crystal attached to the liquid nitrogen pipe by attaching temperature sensors around the side surface of the crystal. However, it is very hard to maintain a good thermal contact between the sensor and the crystal without some mechanism to push the sensor tightly against the surface. Such mechanism would unavoidably introduce extra material around the crystal, which may scatter neutrons in a quenching factor measurement. Instead, we compared light yields of undoped crystals operated in this cryostat and an old cryostat used in our previous light yield measurements to ensure that the crystal was operated at 77~K. The structure of the old crystal is shown in Fig.~\ref{f:old}. As the crystal is fully submerged in liquid nitrogen in this cryostat, its temperature is exactly 77~K. If the light yields measured in the two are similar, the crystals should have been operated at similar temperatures.

Note that the two crystals used in the two cryostats are different in shape, which might create a difference in light collection efficiency. However, because they are all very small, paths of individual photons in them (in the order of centimeters) are much shorter than the average absorption length (in the orders of meters) in CsI. The light collection efficiency hence cannot be too different in the two measurements. 

The detailed description of the old cryostat can be found in Ref.~\cite{ding20e}. An undoped cylindrical crystal with a diameter of 5.08~cm and a height of 1~cm purchased from OKEN~\cite{oken} was used in the old setup. Another difference is the radiation shielding.  As shown in the right Fig.~\ref{f:old}, the old setup has a 1 mm copper in between the $^{241}$Am and the crystal, and two more layers of aluminum foil in between the $^{55}$Fe and the crystal. The sources are directly attached to the surface of the Teflon tape wrapping. 

The same analysis was made on two setups. Only analysis from the new setup is presented here.

\section{Single-photoelectron (SPE) response of the PMT}
\label{s:1pe}
The SPE response of the PMT was measured using light pulses from an ultraviolet LED, LED370E from Thorlabs. Its output spectrum peaked at 375~nm with a width of 10~nm, which was within the 200 -- 650~nm spectral response range of the PMT. Light pulses with a $\sim$50~ns duration and a rate of 10~kHz were generated using a RIGOL DG1022 arbitrary function generator. The intensity of light pulses was tuned by varying the output voltage of the function generator so that only one or zero photons hit the PMT during the LED lit window most of the time. A TTL trigger signal was provided by the function generator simultaneously with each output pulse, then was used to trigger the digitizer to record the PMT response. 

The PMT was biased at 1,600~V, slightly above the recommended operation voltage, 1,500~V, to increase the gain of the PMT.  Single-PE pulses were further amplified by a factor of ten using a Phillips Scientific Quad Bipolar Amplifier Model 771 before being fed into the digitizer in order to separate signals from the pedestal noise.
\begin{figure}[htbp]\centering
  \includegraphics[width=\linewidth]{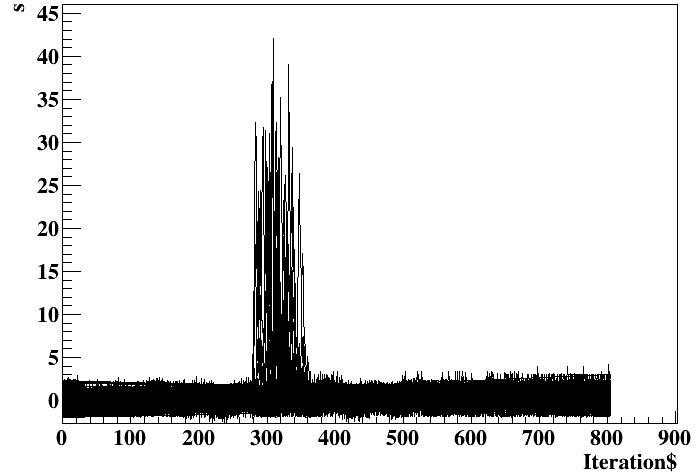}
  \caption{209 selected waveforms from the PMT overlapped with each other.}
  \label{f:1pe}
\end{figure}

The left plot on Fig.~\ref{f:1pe} shows $\sim$210 randomly selected waveforms from the PMT response of the SPE measurement. Some quality criteria were applied to filter out noise events. Firstly, the baseline was calculated in the following steps. The values corresponding to each sample (height) were added to a summed value, then an averaged value (baseline value) was calculated using the summed value divided by the number of samples. The baseline value was then used to shift the waveforms to zero by deducting this baseline value from each corresponding height value. Secondly, the root mean square (RMS) of the baseline was also calculated. After that, averaged baseline values and RMS of the baseline were calculated in a region of 0 to 100 ns, 180 to 280 ns and 400 to 500 ns and all RMS were required to be smaller than 1 ADC count to obtain a relatively stable baseline. The averaged baseline values in the latter two regions were also set to be lower than 1 ADC count to filter out some low-frequency fluctuation events. Another restriction was that the lowest point of the waveform should be greater than $- 2$ ADC counts. Eventually, 209 SPE waveforms were selected applying those restrictions to 1300 SPE events. However, some fluctuations still exist in the pedestal which sits under the SPE waveforms. 
\begin{figure}[htbp]\centering
  \includegraphics[width=\linewidth]{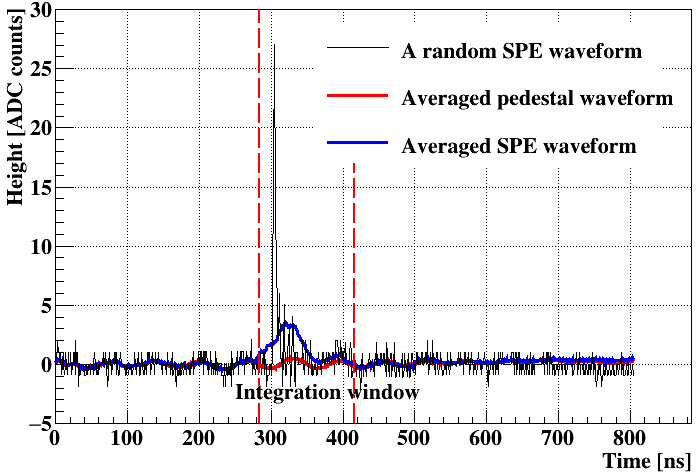}
  \caption{A random SPE waveform, averaged SPE waveform and averaged pedestal waveform.}
  \label{f:ave}
\end{figure}

To examine the influence of the unstable pedestal on the SPE waveforms, the averaged SPE waveforms and the averaged pedestal waveform were studied as shown in the right plot on Fig.~\ref{f:ave}. The averaged SPE waveform is correlated with the averaged pedestal waveform; the baseline and averaged SPE coincide with each other except in the region where SPE initiates. Therefore, the mean area of the SPE can be obtained by subtracting the area of the averaged pedestal waveform from the area of the averaged SPE waveform. The integration window starts from and ends on both pulses crossing zero as shown in the plot. However, SPE waveforms were seen to have much higher heights in the left plot, yet the averaged SPE waveform only has a height of $\sim$5 ADC counts in the right plot. To investigate the discrepancy, we randomly selected an SPE waveform, which was much narrower than the averaged waveform. The wider width of the averaged waveform was due to the fact that the SPE pulses appeared at different locations, ranging from approximately 300 to 360 ns.
\begin{figure}[htbp]\centering
  \includegraphics[width=\linewidth]{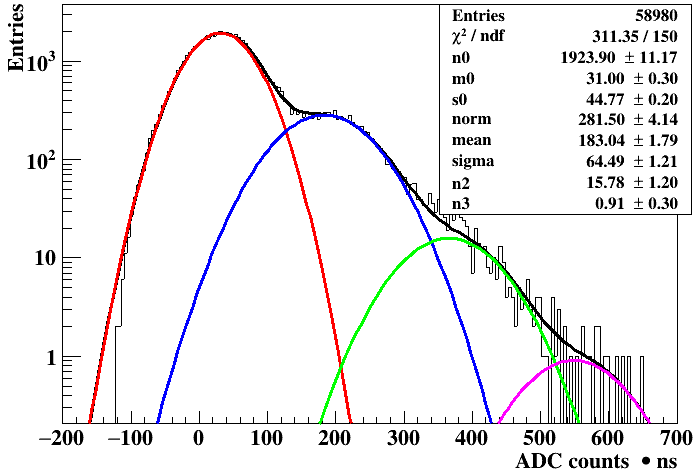}
  \caption{SPE events of the PMT in logarithm scale. The titled peaks are explained in the text.}
  \label{f:spe}
\end{figure}

The integration window (283-416 ns) shown in Fig.~\ref{f:ave} and the same cuts applied to the selected waveforms were then used for each waveform in the data file. The resulting SPE spectrum is shown in Fig.~\ref{f:spe}. The spectrum was fit in the same way as described in Ref.~\cite{ds1013,ding20e}. The red Gaussian curve is fit for the pedestal events, the blue Gaussian curve is fit for the SPE events, and two-PE (green curve) and three-PE fit (pink curve) were Gaussian functions based on the mean and sigma obtained from the SPE fit. The black curve is the sum of those fit results, which matches well the single PE response. The area of single PE (mean$_\text{SPE}$) would be, as mentioned above, mean$_\text{SPE}$ = mean - m$_{0}$, where mean and m$_{0}$ are obtained from the fit result in Fig.~\ref{f:spe}. The value of mean$_\text{SPE}$ is 152.04 ADC counts$\cdot$ns, and the value of m$_{0}$ is 31.00 ADC counts$\cdot$ns. These values will be further used in the light yield calculation. 

To estimate the systematic uncertainty in the determination of the mean value of the SPE distribution, multiple measurements were performed. The discrepancy is within 5\%. In the energy calibration measurements to be mentioned in the next section, the SPE spectrum with the crystal was used but with a 5\% uncertainty attached to be conservative.

\section{Energy calibration}
\label{s:ec}
The energy calibration was performed using an $^{55}$Fe source and an $^{241}$Am source. The digitizer was triggered when the height of a pulse from the PMT was more than 50 ADC counts ($\sim$2 PE). As can be seen in Fig.~\ref{f:1pe}, the height of a single PE pulse was around 25 ADC counts.  The trigger threshold therefore suppresses most of the electronic noise spikes while letting pass most of the PE pulses. The trigger rate was $\sim 2.3$~kHz when the threshold was set to this value.

\begin{figure}[htbp]\centering
    \includegraphics[width=\linewidth]{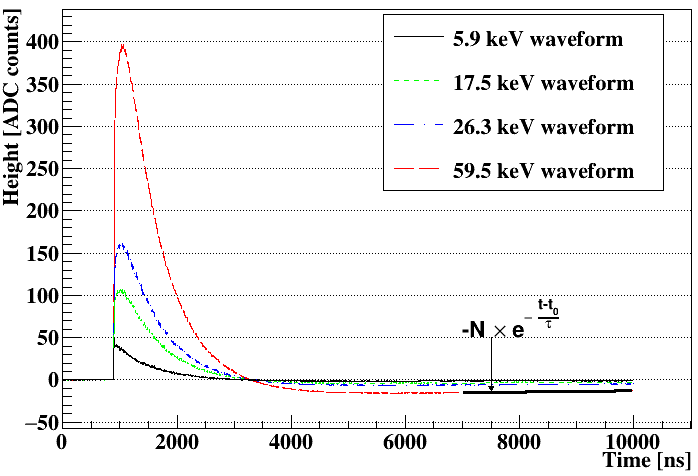}
  \caption{Averaged waveforms with negative overshoot. The tail part of the waveform was fit by an exponential function to estimate the negative overshoot effect.}
  \label{f:Am}
\end{figure}

Each recorded waveform was 10000~ns long as shown in Fig.~\ref{f:Am}. About 1000~ns pre-traces were preserved before the rising edge of a pulse that triggered the digitizer so that there were enough samples before the pulse to calculate the averaged pedestal value of the waveform, and 200 samples starting from zero were used to calculate the baseline. The pedestal was then adjusted to zero using the method described at \autoref{s:1pe}.

By checking a few waveforms, negative overshoot was commonly observed. To further identify whether negative overshoot was common, averaged waveforms were examined. As shown in Fig.~\ref{f:Am}, averaged pulses were computed by summing all the waveforms first, then dividing by the number of events. The tallest one is the averaged 59.5 keV waveform, the second tallest one is the averaged 26.3 keV waveform, the second smallest one is the averaged 17.5 keV waveform, and the smallest one is the averaged 5.9 keV waveform. In the left figure, the negative overshoot effect from the PMT can be clearly seen. 
\begin{figure}[htbp]\centering
    \includegraphics[width=\linewidth]{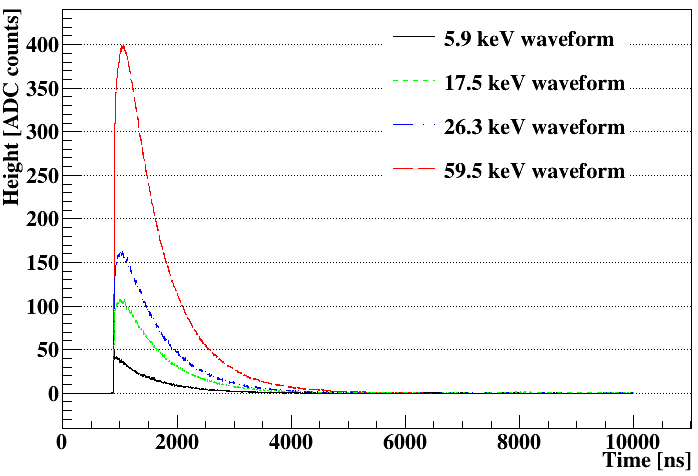}
  \caption{Averaged waveforms after correction.}
  \label{f:crct}
\end{figure}

\subsection{Correction of negative overshoot}
\label{s:eq}
\begin{figure*}[htbp]\centering
  \includegraphics[width=\linewidth]{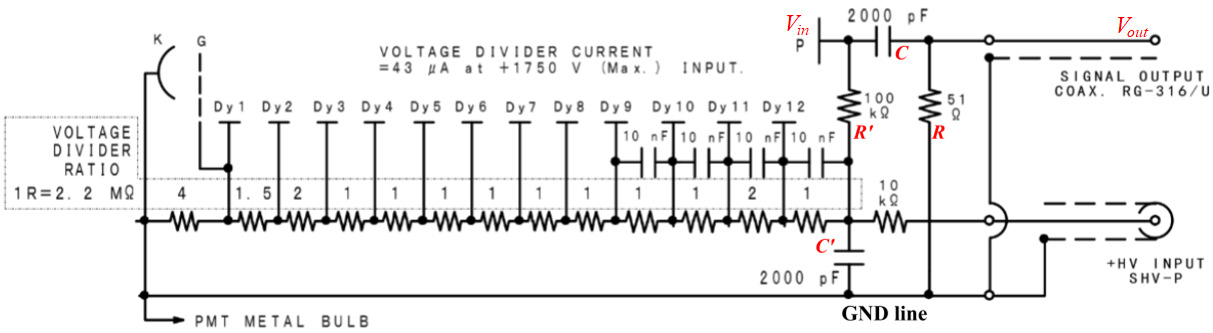}
  \caption{Circuit from Hamamatsu.}
  \label{f:circuit}
\end{figure*}

Fig.~\ref{f:circuit} shows the high voltage (HV) distribution circuit and the readout scheme of our PMT given by its manufacturer, Hamamatsu Photonics K.K. A 2,000 pF capacitor, $C$, is used to decouple the output line from the anode biased at high voltage. The 51 $\Omega$ load resistor, $R$, is used to match the impedance of typical oscilloscopes and digitizers. The waveform of $V_\text{out}$ can be tuned by selecting the values of $C, C', R$, and $R'$ as described in detail in Ref.~\cite{overshoot}. We did not observe any negative overshoot in previous measurements when we used the same PMT. The values of these passive components must have been fine-tuned by Hamamatsu to reduce the negative overshoot. However, their values may have changed over time and through multiple thermal cycles, which resulted in the negative overshoot shown in Fig.~\ref{f:Am}.

As the $RC$ circuit is the origin of the negative overshoot, it is possible to correct this effect offline. A simple numerical method was developed to achieve this. Its derivation is explained step by step here. First, the current going through the load resistor $R$ can be expressed as
\begin{equation}
    I=\frac{V_{out}}{R}=\frac{dQ}{dt},
\end{equation}
where $Q$ is the charge accumulated in $C$. It can be expressed as
\begin{equation}
    Q=C(V_{in}-V_{out}).
\end{equation}
Combing the two equations, we have
\begin{equation}
    \frac{dQ}{dt}=C(\frac{dV_{in}}{dt} - \frac{dV_{out}}{dt} )=\frac{V_{out}}{R},
\end{equation}
which can be rearranged as
\begin{equation}
    V_\text{out} = RC\left(\frac{dV_\text{in}}{dt}-\frac{dV_\text{out}}{dt}\right)
\end{equation}
Numerically, this can be written as
\begin{equation}\label{e:vout}
 V_{out}[i]=RC\left(\frac{V_{in}[i]-V_{in}[i-1]}{\Delta t}-\frac{V_{out}[i]-V_{out}[i-1]}{\Delta t}\right),
\end{equation}
where $i$ is the index of individual samples in the waveform, and $\Delta t = 1$~ns is the time interval between two consecutive samples. The iterative expression of $V_\text{in}$ can be derived from Eq.~\ref{e:vout}:
\begin{equation} \label{e:vin}
     V_{in}[i]=\frac{RC+\Delta t}{RC}V_{out}[i]-V_{out}[i-1]+V_{in}[i-1].
\end{equation}
The $RC$ constant in Eq.~\ref{e:vin} was measured to be $18,044$~ns by fitting a simple exponential function to the averaged waveforms as shown in Fig.~\ref{f:Am} in the range of $[7,000, 10,000]$ ns, where the influence of scintillation decay can be neglected and the influence of the $RC$ circuit persists. 

Therefore, V$_{out}$ can be corrected sample by sample once the RC constant was identified. The averaged waveforms after correction shown in Fig.~\ref{f:crct} demonstrate the success of this correction method and the reasonableness of the RC constant. The same correction method with the same RC constant was then applied to each waveform. This method produces reasonable waveforms and doesn't make any assumptions on the input signal.

\begin{figure}[htbp]\centering
  \includegraphics[width=\linewidth]{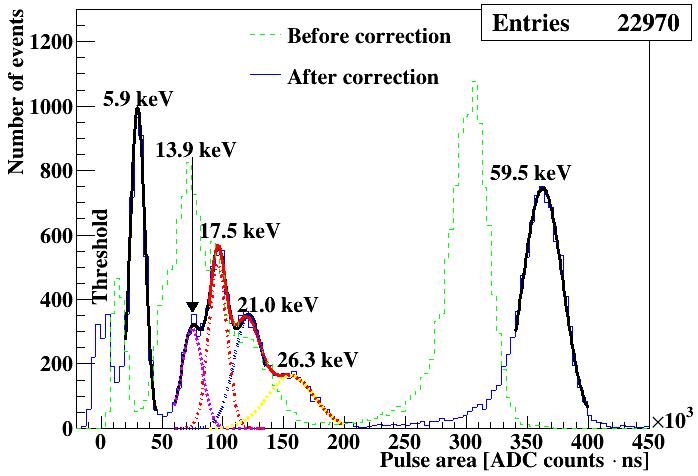}
  \caption{Energy spectra of $^{55}$Fe and $^{241}$Am in units of ADC counts$\cdot$ns.}
  \label{f:ec}
\end{figure}

Every waveform after the correction was integrated from zero to the end. The integration had a unit of ADC counts$\cdot$ns. In Fig.~\ref{f:ec}, the recorded energy spectra before correction are the green dashed line histogram, and the energy spectra after correction are the blue solid line histogram. A dominant 5.9 keV peak from $^{55}$Fe can be seen, 17.5 keV, 26.3 keV and 59.5 keV peaks from $^{241}$Am are shown as well. Fits were applied to the corrected spectra. The Gaussian fit was applied to the 5.9 and 59.5 keV peaks. The combined fit was applied to 13.9, 17.5 and 21.0 keV peaks, and then applied to 17.5, 21.0 and 26.3 keV peaks. Pulse area means obtained from those fits are summarized in~\autoref{t:rPE}, and whether the light response stays consistent in the energy close to the threshold is evaluated in the next section. 

\section{Light yield}
\label{s:ly}
The pulse area values ($A$) of the radiation pulses in the unit of ADC counts$\cdot$ns were converted to the number of PE ($N_\text{PE}$) using the formula:
\begin{equation}
 N_\text{PE} = \frac{A-m_0}{\text{mean}_\text{SPE}}.
  \label{e:m1pe}
\end{equation}

The shift value, $m_{0}$, is added to account for the overall shift of the pulses observed in the single PE measurement. However, compared to the pulse area values, the shift is small.

The light yield ($Y$) for a given energy deposited ($E_\text{dep}$) in our electron recoil measurements was then calculated using the following equation:
\begin{equation}
  Y [\text{PE/keV}_\text{ee}] = \frac{N_\text{PE}}{E_\text{dep}}.
  \label{e:ly}
\end{equation}

\begin{table*}[htbp]
  \caption{\label{t:rPE} Fit results along with the calculated light yield of $^{59}$Fe and $^{241}$Am peaks in the energy spectrum are shown. The top table is from the new cryostat, whereas the bottom table is from the old cryostat.}
  \begin{minipage}{\linewidth}\centering
  \begin{tabular}{l @{\extracolsep{\fill}}ccccccc}
    \toprule
    Type of & Energy & Mean (A) & Sigma & FWHM & Light yield & Uncertainty\\
   radiation & [keV] & [ADC$\cdot$ns] & [ADC$\cdot$ns] & \% & [PE/keV$_\text{ee}$] & [PE/keV$_\text{ee}$] \\
    \midrule
     \begin{tabular}{r} x-ray\\ x-ray\\ x-ray \\ x-ray \\ $\gamma$-ray \\ $\gamma$-ray \\
     \end{tabular} &
     \begin{tabular}{l} 5.9 \\ 13.9$^\dagger$ \\ 17.5$^\dagger$ \\ 21.0$^\dagger$ \\ 26.3$^\dagger$ \\ 59.5 \\
     \end{tabular} &
     \begin{tabular}{r} 29984 \\ 74709 \\ 95885 \\ 121409\\ 155757 \\ 362693 \\
     \end{tabular} &
     \begin{tabular}{r} 6138 \\ 8330 \\ 7516 \\ 12000 \\ 19322 \\ 16607 \\
     \end{tabular} &
     \begin{tabular}{r} 48.2 \\ 26.3 \\ 17.8 \\ 23.3 \\ 29.2 \\ 10.8 \\
     \end{tabular} &
     \begin{tabular}{r} 33.4 \\ 35.3 \\ 36.0 \\ 38.0 \\ 38.9 \\ 40.0 \\
     \end{tabular} &
    \begin{tabular}{r} $\pm$ 1.7 \\ $\pm$ 1.8 \\ $\pm$ 1.8 \\ $\pm$ 1.9 \\ $\pm$ 1.9 \\ $\pm$ 2.0 \\
     \end{tabular} \\
  \bottomrule
     \end{tabular}
  \end{minipage}
  
  \begin{minipage}{\linewidth}\centering
  \begin{tabular}{l @{\extracolsep{\fill}}ccccccc}
    \toprule
    Type of & Energy & Mean (A) & Sigma & FWHM & Light yield & Uncertainty\\
   radiation & [keV] & [ADC$\cdot$ns] & [ADC$\cdot$ns] & \% & [PE/keV$_\text{ee}$] & [PE/keV$_\text{ee}$] \\
    \midrule
     \begin{tabular}{r} x-ray\\ $\gamma$-ray \\ $\gamma$-ray \\
     \end{tabular} &
     \begin{tabular}{l} 5.9 \\ 26.3 \\ 59.5 \\
     \end{tabular} &
     \begin{tabular}{r} 32261 \\ 166507 \\ 354948 \\
     \end{tabular} &
     \begin{tabular}{r} 5454 \\ 14310 \\ 13264 \\
     \end{tabular} &
     \begin{tabular}{r} 39.8 \\ 20.2 \\ 8.8 \\
     \end{tabular} &
     \begin{tabular}{r} 35.9 \\ 41.6 \\ 39.2\\
     \end{tabular} &
     \begin{tabular}{r} $\pm$ 1.8 \\ $\pm$ 2.1 \\ $\pm$ 2.0\\
     \end{tabular} \\
     \bottomrule
     \end{tabular}
  \end{minipage}
  $^\dagger$ Intensity averaged mean of x-rays near each other~\cite{ding20e}.\\
\end{table*}

The calculated light yields in different setups are shown in \autoref{t:rPE}, similar light yields observed from two cryostats confirm the cooling capability of the new cryostat. 

In the new cryostat, the light yield from the $^{59}$Fe 5.9 keV radiation is 33.4 $\pm$ 1.7 PE$/$keV$_\text{ee}$, while the light yield from the $^{241}$Am 59.5 keV radiation is 40.0 $\pm$ 2.0 PE$/$keV$_\text{ee}$. The variation in the light yield at different energies is also seen in the old cryostat setup, with a light yield of 35.9 $\pm$ 1.8 PE$/$keV$_\text{ee}$ from the 5.9 keV radiation and a light yield of 39.2 $\pm$ 2.0 PE$/$keV$_\text{ee}$ from the 59.5 keV radiation.

The observed nonlinearity of pure CsI crystal at different energies was also seen in Ref.~\cite{csi, csi20, ding20e, ding20, ding22, LY_moszynski_energy_2005, LY_gridin_channels_2014, LY_kerisit_computer_2009}.  Crystals from different vendors show slightly different behavior. Light yield decreases slightly as energy goes down. One possible explanation is that low-energy radiation cannot penetrate far into the crystal, the survivability of optical photons created around the surface of the crystal depends highly on the local surface condition; while high-energy radiation on average creates optical photons deep inside the crystal, their survivability does not depend on the surface condition. 

\section{Conclusion}

This study describes the development of a liquid nitrogen-based cryostat setup for measuring the nuclear quenching factor of undoped CsI crystals at cryogenic temperatures. The cryostat was designed to avoid neutron scattering in high-$Z$ materials and achieved a light yield of $33.4 \pm 1.7$ PE$/$keV$_\text{ee}$ at 5.9 keV$_\text{ee}$. The results of the quenching factor measurement will be reported in a subsequent paper. Additionally, the study highlights the issue of non-negligible negative overshoot in the tails of light pulses and provides a detailed explanation of its origin and correction procedure. The information presented in this study will be valuable for others working in similar fields and encountering similar technical challenges.

\section*{Acknowledgements}
This work is supported by the Department of Energy (DOE), USA, award DE-SC0022167, and the National Science Foundation (NSF), USA, award PHY-1506036. Computations supporting this project were performed on High Performance Computing systems at the University of South Dakota, funded by NSF award OAC-1626516.

\bibliographystyle{achemso}
\bibliography{ref}
\end{document}